\documentclass[final,5p,times,twocolumn]{elsarticle}

\usepackage{amssymb}
\usepackage{hyperref}
\usepackage{booktabs}

\usepackage{lineno}
\usepackage{comment}
\usepackage{caption}
\usepackage{subcaption}

\journal{Nuclear Inst. and Methods in Physics Research, A}

\begin{document}

\begin{frontmatter}

\title{Exploring Eco-Friendly Gas Mixtures for Resistive Plate Chambers: A Comprehensive Study on Performance and Aging}

\author[1]{\small \textbf{The RPC ECOgas@GIF++ Collaboration: }L. Quaglia} 
\author[6,3]{\small M. Abbrescia}
\author[21]{\small G. Aielli}
\author[3,8]{\small R. Aly}
\author[17]{\small M. C. Arena}
\author[11]{\small M. Barroso}
\author[4]{\small L. Benussi}
\author[4]{\small S. Bianco}
\author[7]{\small D. Boscherini}
\author[16]{\small F. Bordon}
\author[7]{\small A. Bruni}
\author[18]{\small S. Buontempo}
\author[16]{\small M. Busato}
\author[21]{\small P. Camarri}
\author[13]{\small R. Cardarelli}
\author[3]{\small L. Congedo}
\author[11]{\small D. De Jesus Damiao}
\author[6,3]{\small M. De Serio}
\author[21]{\small A. Di Ciaccio}
\author[21]{\small L. Di Stante}
\author[14]{\small P. Dupieux}
\author[19]{\small J. Eysermans}
\author[12,1]{\small A. Ferretti}
\author[6,3]{\small G. Galati}
\author[12,1]{\small M. Gagliardi}
\author[16]{\small R. Guida}
\author[2,3]{\small G. Iaselli}
\author[14]{\small B. Joly}
\author[24]{\small S.A. Juks}
\author[23]{\small K.S. Lee}
\author[13]{\small B. Liberti} 
\author[22]{\small D. Lucero Ramirez}
\author[16]{\small B. Mandelli}
\author[14]{\small S.P. Manen}
\author[7]{\small L. Massa}
\author[3]{\small A. Pastore}
\author[13]{\small E. Pastori}
\author[4]{\small D. Piccolo}
\author[13]{\small L. Pizzimento}
\author[7]{\small A. Polini}
\author[13]{\small G. Proto}
\author[2,3]{\small G. Pugliese}
\author[2,3]{\small D. Ramos}
\author[16]{\small G. Rigoletti}
\author[13]{\small A. Rocchi}
\author[7]{\small M. Romano}
\author[10]{\small A. Samalan}
\author[9]{\small P. Salvini}
\author[21]{\small R. Santonico}
\author[5]{\small G. Saviano}
\author[13]{\small M. Sessa}
\author[6,3]{\small S. Simone}
\author[12,1]{\small L. Terlizzi}
\author[10,20]{\small M. Tytgat}
\author[12,1]{\small E. Vercellin}
\author[15]{\small M. Verzeroli}
\author[22]{\small N. Zaganidis}

\affiliation[1]{organization={INFN Sezione di Torino},
             addressline={Via P. Giuria 1},
             city={Torino},
             postcode={10125},
             state={Italy},
             country={}}
             
\affiliation[2]{organization={Politecnico di Bari, Dipartimento Interateneo di Fisica},
             addressline={via Amendola 173},
             city={Bari},
             postcode={70125},
             state={Italy},
             country={}}

\affiliation[3]{organization={INFN Sezione di Bari},
             addressline={Via E. Orabona 4},
             city={Bari},
             postcode={70125},
             state={Italy},
             country={}}

\affiliation[4]{organization={INFN - Laboratori Nazionali di Frascati},
             addressline={Via Enrico Fermi 54},
             city={Frascati (Roma)},
             postcode={00044},
             state={Italy},
             country={}}

\affiliation[5]{organization={Sapienza Università di Roma, Dipartimento di Ingegneria Chimica Materiali Ambiente},
             addressline={Piazzale Aldo Moro 5},
             city={Roma},
             postcode={00185},
             state={Italy},
             country={}}

\affiliation[6]{organization={Università degli studi di Bari, Dipartimento Interateneo di Fisica},
             addressline={Via Amendola 173},
             city={Bari},
             postcode={70125},
             state={Italy},
             country={}}

\affiliation[7]{organization={INFN Sezione di Bologna},
             addressline={Via C. Berti Pichat 4/2},
             city={Bologna},
             postcode={40127},
             state={Italy},
             country={}}

\affiliation[8]{organization={Helwan University},
             addressline={},
             city={Helwan, Cairo Governorate},
             postcode={4037120},
             state={Egypt},
             country={}}

\affiliation[9]{organization={INFN Sezione di Pavia},
             addressline={Via A. Bassi 6},
             city={Pavia},
             postcode={27100},
             state={Italy},
             country={}}

\affiliation[10]{organization={Ghent University, Dept. of Physics and Astronomy},
             addressline={Proeftuinstraat 86},
             city={Ghent},
             postcode={B-9000},
             state={Belgium},
             country={}} 
             
\affiliation[11]{organization={Universidade do Estado do Rio de Janeiro},
             addressline={R. São Francisco Xavier, 524},
             city={Maracanã, Rio de Janeiro - RJ},
             postcode={20550-013},
             state={Brazil},
             country={}} 

\affiliation[12]{organization={Università degli studi di Torino, Dipartimento di Fisica},
             addressline={Via P. Giuria 1},
             city={Torino},
             postcode={10125},
             state={Italy},
             country={}} 

\affiliation[13]{organization={INFN Sezione di Roma Tor Vergata},
             addressline={Via della Ricerca Scientifica 1},
             city={Roma},
             postcode={00133},
             state={Italy},
             country={}}

\affiliation[14]{organization={Clermont Université, Université Blaise Pascal, CNRS/IN2P3, Laboratoire de Physique Corpusculaire},
             addressline={BP 10448},
             city={Clermont-Ferrand},
             postcode={F-63000},
             state={France},
             country={}}

\affiliation[15]{organization={Universitè Claude Bernard Lyon I},
             addressline={43 Bd du 11 Novembre 1918},
             city={Villeurbanne},
             postcode={69100},
             state={France},
             country={}}

 \affiliation[16]{organization={CERN},
             addressline={Espl. des Particules 1},
             city={Meyrin},
             postcode={1211},
             state={Switzerland},
             country={}}   
             
\affiliation[17]{organization={Università degli studi di Pavia},
             addressline={Corso Strada Nuova 65},
             city={Pavia},
             postcode={27100},
             state={Italy},
             country={}} 

\affiliation[18]{organization={INFN Sezione di Napoli},
             addressline={Complesso universitario di Monte S. Angelo ed. 6 Via Cintia},
             city={Napoli},
             postcode={80126},
             state={Italy},
             country={}}

\affiliation[19]{organization={Massachusetts Institute of Technology},
             addressline={77 Massachusetts Ave},
             city={Cambridge, MA},
             postcode={02139},
             state={USA},
             country={}}

\affiliation[20]{organization={Vrije Universiteit Brussel (VUB-ELEM), Dept. of Physics},
             addressline={Pleinlaan 2},
             city={Brussels},
             postcode={1050},
             state={Belgium},
             country={}}

\affiliation[21]{organization={Università degli studi di Roma Tor Vergata, Dipartimento di Fisica},
             addressline={Via della Ricerca Scientifica 1},
             city={Roma},
             postcode={00133},
             state={Italy},
             country={}}

\affiliation[22]{organization={Universidad Iberoamericana, Dept. de Fisica y Matematicas},
             addressline={},
             city={Mexico City},
             postcode={01210},
             state={Mexico},
             country={}}

\affiliation[23]{organization={Korea University},
             addressline={145 Anam-ro},
             city={Seongbuk-gu, Seoul},
             postcode={},
             state={Korea},
             country={}}

\affiliation[24]{organization={Université Paris-Saclay},
             addressline={3 rue Joliot Curie, Bâtiment Breguet},
             city={Gif-sur-Yvette},
             postcode={91190},
             state={France},
             country={}}

\begin{abstract}

\small Resistive Plate Chambers (RPCs) are gaseous detectors widely used in high energy physics experiments, operating with a gas mixture primarily containing Tetrafluoroethane (C$_{2}$H$_{2}$F$_{4}$), commonly known as R-134a, which has a global warming potential (GWP) of 1430. To comply with European regulations and explore environmentally friendly alternatives, the RPC EcoGas@GIF++ collaboration, involving ALICE, ATLAS, CMS, LHCb/SHiP, and EP-DT communities, has undertaken intensive R\&D efforts to explore new gas mixtures for RPC technology. 

\small A leading alternative under investigation is HFO1234ze, boasting a low GWP of 6 and demonstrating reasonable performance compared to R-134a. Over the past few years, RPC detectors with slightly different characteristics and electronics have been studied using HFO and CO$_{2}$-based gas mixtures at the CERN Gamma Irradiation Facility. An aging test campaign was launched in August 2022, and during the latest test beam in July 2023, all detector systems underwent evaluation. This contribution will report the results of the aging studies and the performance evaluations of the detectors with and without irradiation.

\end{abstract}


\end{frontmatter}

\section{Introduction}
\label{sec:intro}

Resistive Plate Chambers (RPCs) are gaseous detectors with planar geometry and resistive electrodes (either made out of bakelite or glass). Thanks to their relatively low-cost and $\approx$ns time resolution they are widely employed in the muon trigger/identification systems of the LHC experiments \cite{muonTDR,cmsTDR,atlasTDR}. These RPCs are operated in avalanche mode, with a gas mixture containing a high fraction ($>$~90\%) of C$_{2}$H$_{2}$F$_{4}$ and SF$_{6}$ (plus a fraction of i-C$_{4}$H$_{10}$ as photon quencher) and, although this mixture satisfies all the performance requirements, it contains a high fraction of C$_{2}$H$_{2}$F$_{4}$ and SF$_{6}$, which are classified as fluorinated greenhouse gases (F-gases/GHGs).

Starting from 2014, new European Union regulations \cite{euReg} have imposed a progressive phase-down in the production and usage of these compounds, leading to an increase of cost and reduction in availability. For this reason, CERN has adopted a policy of F-gases reduction and, since RPCs represent a significant fraction of the total GHG-gases emission of the LHC experiments\cite{beatrice}, it is of the utmost importance to search for more eco-friendly RPC gas mixtures. The first efforts have been concentrated on the replacement of C$_{2}$H$_{2}$F$_{4}$ using its industrial replacement, the \textit{tetrafluoropropene} (C$_{3}$H$_{2}$F$_{4}$ or simply HFO, in its -ze isomer) diluted with other gases to lower the detector working voltage \cite{prelGiorgia,prelAntonio,prelGianluca,prelPiccolo}. Promising gas mixtures, where C$_{2}$H$_{2}$F$_{4}$ has been replaced with different fractions of HFO/CO$_{2}$ have been identified and now a more complete characterization of those mixtures in controlled data-taking environments (such as beam tests), as well as the study of the detectors long-term behavior (aging studies) when operated with these new gas mixtures is needed.

To this aim, the RPC ECOgas@GIF++ collaboration (including researchers from ALICE, ATLAS, CMS, LHCb/SHiP and the CERN EP-DT group) was created to join forces among RPC experts of the different LHC experiments, sharing knowledge and manpower. Each group has provided a detector prototype, which has been installed on a common mechanical support and both beam tests as well as aging studies are being carried out using HFO-based gas mixtures. This contribution will provide an overview of the latest results obtained from the ongoing aging studies.

The text is divided as follows: Section \ref{sec:setup} contains a description of the experimental setup and the methodology used in the data-taking/analysis, Section \ref{sec:results} reports the main results obtained from the ongoing aging campaign where mixtures with different HFO/CO$_{2}$ ratios are being studied and, lastly, Section \ref{sec:conclusion} is dedicated to the conclusion and to possible outlooks for the future of this work.

\section{Experimental setup and methodology}
\label{sec:setup}

The experimental setup of the RPC ECOgas@GIF++ collaboration is installed at the Gamma Irradiation Facility (GIF++)\cite{GIF++}, located on the H4 secondary SPS beam line at CERN. This facility is equipped with a high activity $^{137}$Cs source ($\approx$12.5~TBq), which can be used to induce a high background radiation on the detectors under test, allowing one to simulate long operation periods in a much shorter time-span (years in $\approx$months). The radiation from the $^{137}$Cs source can be modulated by means of a set of 3$\times$3 lead attenuation filters, leading to 27 possible irradiation intensities. Moreover, in dedicated periods, the facility is also traversed by a high-energy (150~GeV) muon beam, allowing one to periodically test the performance of the detectors under study. 

Table \ref{tab:detectors} reports the main features of each detector of the RPC ECOgas@GIF++ collaboration.  Note that the CMS RE11 RPC is the only one having a double gas gap: top (T in Table \ref{tab:detectors}, divided in two: top-wide (TW)) and bottom (B in Table \ref{tab:detectors} or BOT for short). Moreover, it is the only one with a trapezoidal shape while all the others are rectangles.

\begin{table}[h!]
	\begin{center}
		\setlength{\tabcolsep}{1.3pt} 
        \begin{tabular}{ccccc}\\\toprule  
        \textbf{Name} & \textbf{Gaps} & \textbf{Gap (mm)} & \textbf{Electrode (mm)} & \textbf{Area (cm$^{2}$)} \\\midrule
        ALICE & 1 & 2 &  2 & 2500 \\  \midrule
        ATLAS & 1 & 2 & 2 & 550  \\  \midrule
        CMS RE11 & 2 & 2 & 2 & 3627(T)/4215(B)\\  \midrule
        EP-DT & 1 & 2 & 2 & 7000 \\  \midrule
        LHCb/SHiP & 1 & 1.6 & 1.6 & 7000 \\  \bottomrule
        \end{tabular}
        \caption{Features of the RPC ECOgas@GIF++ collaboration detectors}\label{tab:detectors}
	\end{center}
\end{table}

\begin{figure}[h!]
\includegraphics[width=\linewidth]{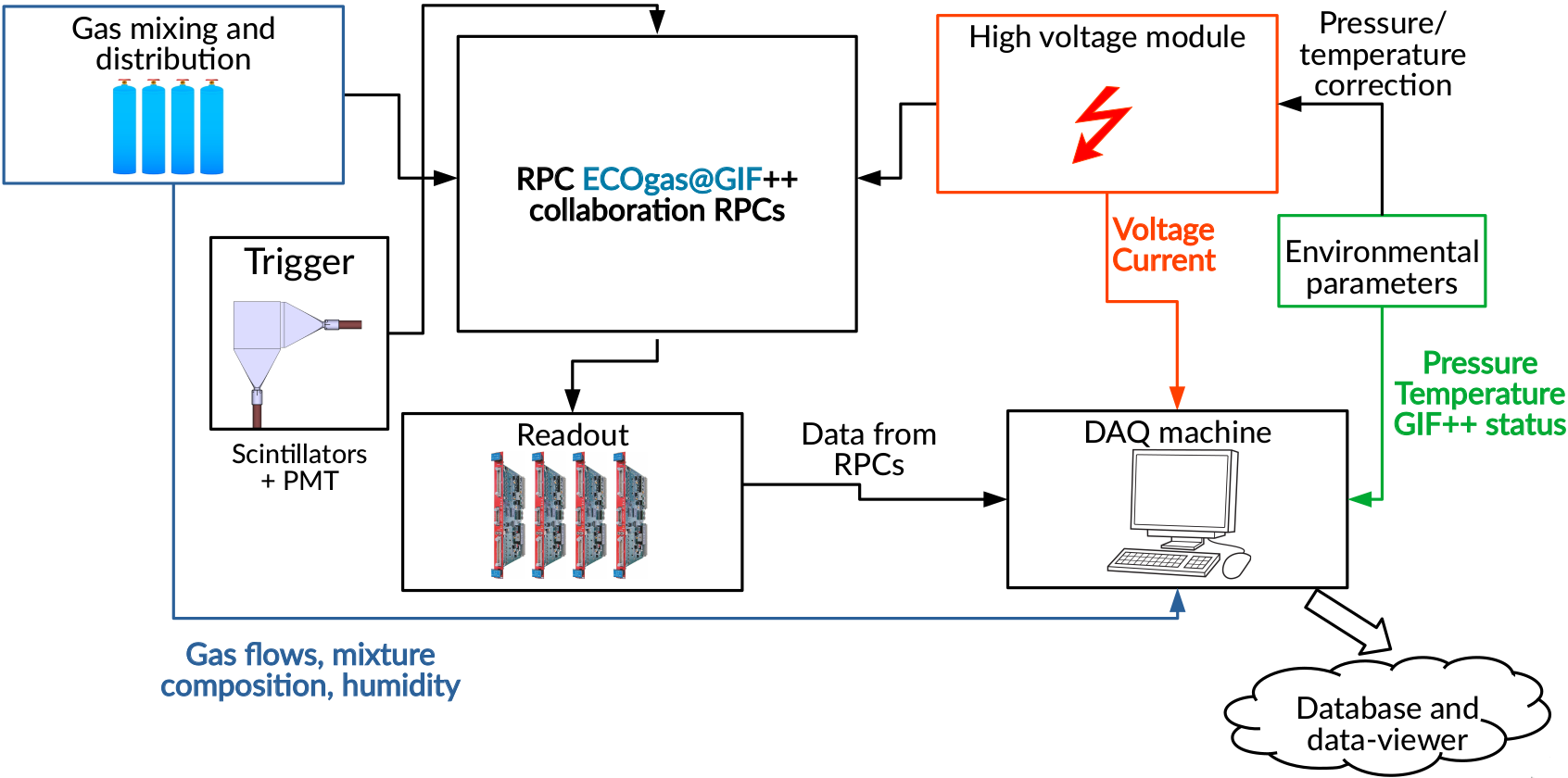}
\caption{Sketch of the experimental setup installed in GIF++ (a detailed description of each component is reported in the text)}
\label{fig:setup}
\end{figure}

Figure \ref{fig:setup} shows a sketch of the experimental setup currently installed at the GIF++. The gas mixing and distribution system allows the users to mix up to four gases, by means of the same amount of mass flow controllers (MFCs). Two more MFCs are used to regulate the relative humidity of the mixture by changing the amount of gas flowing in a humidifier tank. The high voltage (HV) to power the detectors is provided by means of a CAEN SY1527 mainframe\footnote{\url{https://www.caen.it/subfamilies/mainframes/}} and two CAEN A1526 boards\footnote{\url{https://www.caen.it/products/a1526/}} (one with positive and one with negative polarity). A monitoring software is employed to continuously store all the relevant parameters (i.e. environmental conditions, gas mixture composition as well as current absorbed by the detectors and applied high voltage) on a dedicated database for later analysis. During the beam test campaigns, a scintillator-based trigger system is also installed and the data from the detectors are acquired by the readout system (which is detector specific, as detailed in \cite{focusPoint}). For what concerns the aging studies, a dedicated software \cite{webdcs} (referred to as \textit{webdcs}), specifically developed for the needs of the CMS collaboration studies at the GIF++, has been adapted for the RPC ECOgas@GIF++ collaboration studies and it is employed as an online Detector Control System (DCS) for the studies described in the following.  

\subsection{Methodology}
\label{sub:methodology}

During the ongoing aging studies (which began in July 2022), the stability in time of the absorbed current has been studies. In particular, the HV applied to the detectors is set at a fixed value of effective high voltage (i.e. corrected for temperature and pressure variations) and the absorbed current is sampled with a frequency of 1 measurement  every 30~s. 

\begin{figure}[h!]
\center \includegraphics[height=0.5\linewidth]{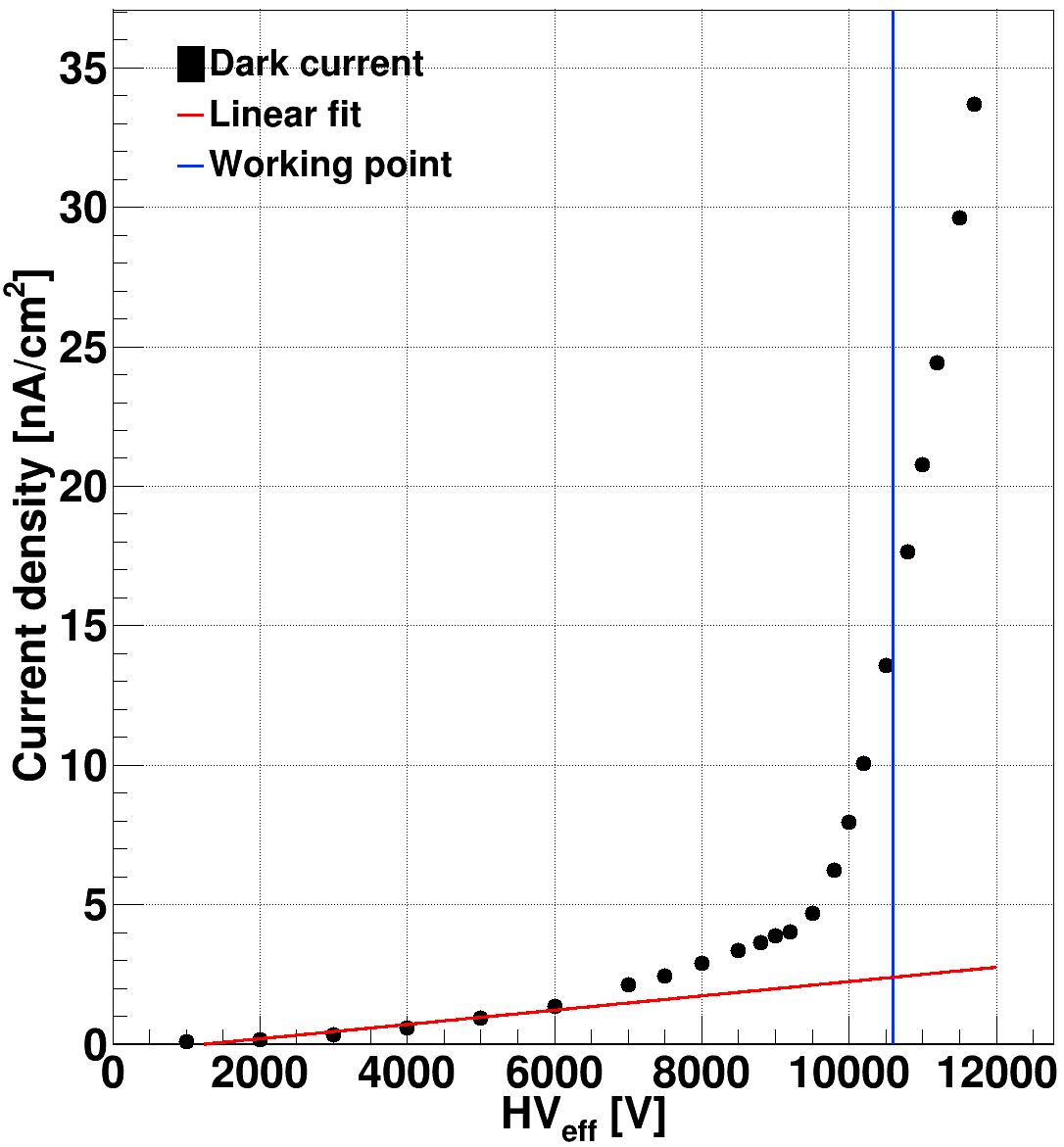}
\caption{Example of dark current scan, obtained from the EP-DT detector. The linear interpolation to estimate the Ohmic dark current at the irradiation voltage is also shown in red in the figure and the blue line represents the irradiation voltage}
\label{fig:darkEx}
\end{figure}

Moreover, once a week, the $^{137}$Cs source is fully shielded (source-off in the following) and a measurement of the current absorbed without any background is carried out as a function of the high voltage (dark current scan). The aim of this operation is two-fold: on the one hand it is used to monitor the stability of the dark current over time (an increase of this quantity could be a sign of potential detector aging) and, on the other, it is used to more accurately estimate the charge integrated by the RPCs during the aging. Indeed, during aging studies, the integrated charge (defined as the integral of the current in time) is used as a proxy for the \textit{age} of the detector. Figure \ref{fig:darkEx} shows an example of a dark current scan from the EP-DT detector. One can see that, for voltages below the gas multiplication threshold ($\approx$7-8~kV for a 2~mm gas gap detector), a non-zero current is flowing through the detector and, given the voltage range, this current is not passing through the gas but rather through other conductive paths in the RPCs and, therefor, it is not directly contributing to the gas-induced aging of the detectors. By looking at Figure \ref{fig:darkEx}, one can see that in the range 0-5~kV the current increases linearly (for this reason this is usually referred to as \textit{Ohmic dark current}) with the HV and, by carrying out a linear interpolation in this range, one can estimate the Ohmic dark current at the irradiation voltage (working point) and subtract it from the total current absorbed under irradiation, to get a more realistic estimate of the actual current flowing through the gas and use this to calculate the integrated charge.

\vspace{-25pt}

\section{Results}
\label{sec:results}

Two mixtures have been aging-tested by the RPC ECOgas@GIF++ collaboration (referred to as ECO1 and ECO2 in the following). Their composition is reported in Table \ref{tab:mixtures}, together with the currently employed gas mixture (STD), which is taken as a reference to which the eco-friendly alternatives have to be compared.

\begin{table}[h!]
	\begin{center}
		\setlength{\tabcolsep}{1.1pt} 
        \begin{tabular}{cccccc}\\\toprule  
        \textbf{Name} & \textbf{C$_{2}$H$_{2}$F$_{4}$ (\%)} & \textbf{HFO (\%)} & \textbf{CO$_{2}$ (\%)} & \textbf{i-C$_{4}$H$_{10}$ (\%)} & \textbf{SF$_{6}$ (\%)} \\\midrule
        STD & 95.2 & 0 &  0 & 4.5 & 0.3 \\  \midrule
        ECO1 & 0 & 45 & 50 & 4 & 1\\  \midrule
        ECO2 & 0 & 35 & 60 & 4 & 1\\  \bottomrule
        \end{tabular}
        \caption{Composition of the gas mixtures used in the aging studies}\label{tab:mixtures}
	\end{center}
\end{table}

Some preliminary studies using the STD gas mixture, described for example in \cite{phdLuca}, have shown that the data-taking procedure and apparatus work as expected hence a first eco-friendly alternative was tested (ECO1). The working point of this mixture was found to be $\approx$2~kV higher than the STD gas mixture (11.6~kV rather than 9.6~kV for the 2~mm gap detectors). Moreover, it was observed that, given the same background radiation level, the current absorbed by the detectors operated with the ECO1 mixture, was $\approx$1.7 times higher than with the STD gas mixture. Following the integration of $\approx$10~mC/cm$^{2}$ all the detectors of the collaboration started to show signs of current instability (both in the dark as well as in the one absorbed under irradiation); moreover, the greater absorbed current could also be related to a higher impurity production under irradiation. These two observations lead to discard this mixture from further studies and the ECO2 mixture, with a reduced HFO content (and a lower working point) was introduced. Using this mixture the currents were still higher than with the STD one (by a factor $\approx$1.5) but the detectors behaved in a more stable manner and a longer irradiation campaign was launched. 

The latter started in July 2022 (concurrently with a beam test, used to set the baseline performance of the RPCs) and is still ongoing at the time of writing. For most of the irradiation period, the background radiation has been $\approx$500~Hz/cm$^{2}$ and, to limit the current absorbed by the detectors, it was decided to operate them at $\approx$70\% efficiency. 

\begin{figure}[h!]
\center \includegraphics[height=0.6\linewidth]{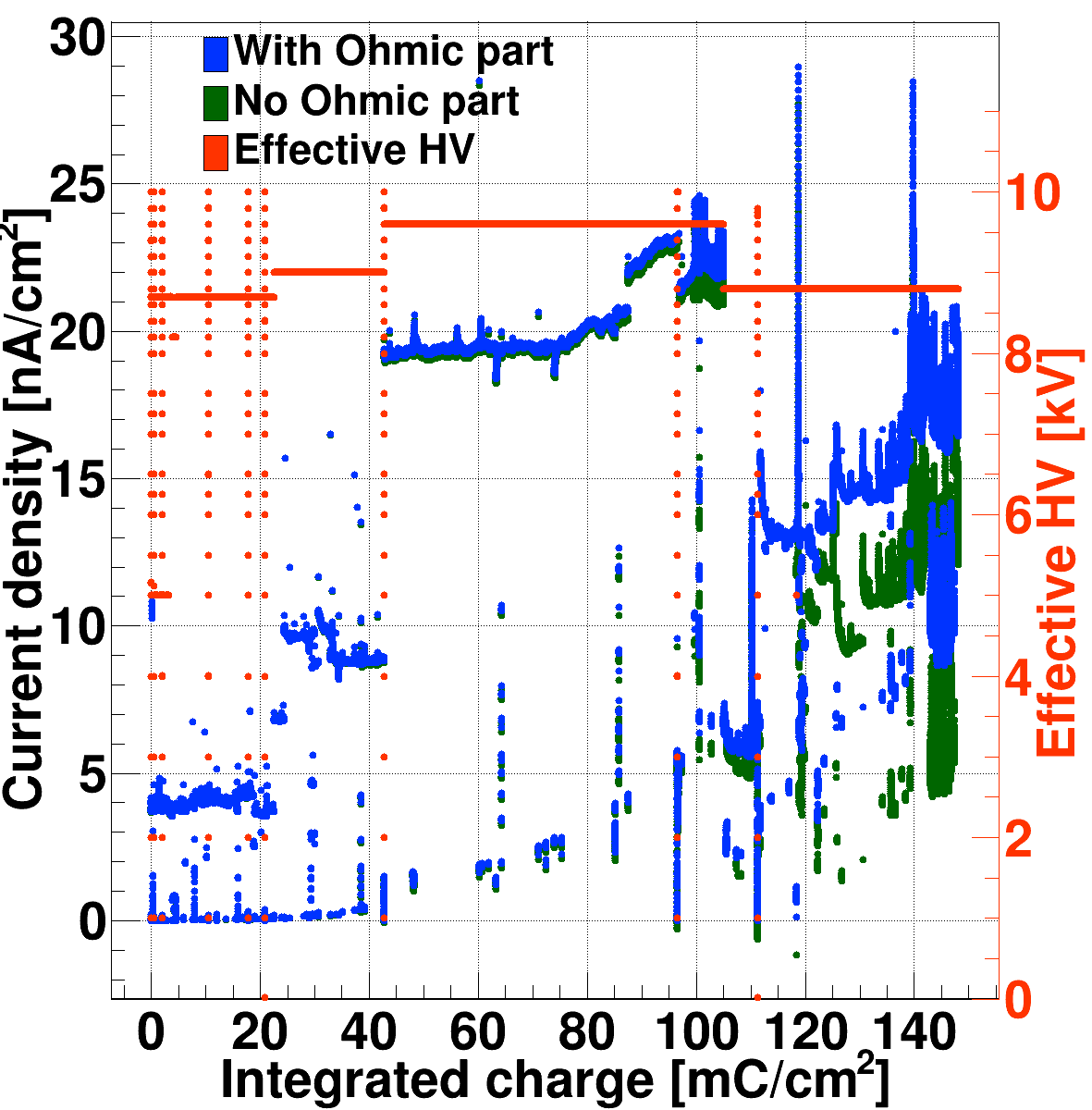}
\caption{Trend of current and effective high voltage as a function of the integrated charge for the LHCb/SHiP RPC}
\label{fig:agingShip}
\end{figure}

Figure \ref{fig:agingShip} shows the main results obtained during the aging studies, for the LHCb/SHiP detector. Rather than as a function of time, the data are shown as a function of the integrated charge (calculated with the subtraction of the Ohmic dark current, as explained in Section \ref{sec:setup}). The data-points in blue in Figure \ref{fig:agingShip} represent the total current density (in nA/cm$^{2}$) absorbed by detector (as measured by the HV module) while the ones in green represent the total current density with the subtraction of the Ohmic dark current (following the procedure highlighted in Section \ref{sec:setup}). 

It has to be noted that the current is shown independently of the irradiation condition and the portions of the figure where the current is lower then 5 nA/cm$^{2}$ represent periods of time without irradiation while changing HV values correspond to the dark current scans mentioned earlier. 

The HV applied to the LHCb/SHiP RPC has been increased in steps (from 8.5 up to 9.8~kV), being the only detector of the collaboration to reach full efficiency under irradiation. By looking at Figure \ref{fig:agingShip} a few comments can be made: in the 0-20~mC/cm$^{2}$ range both the current under irradiation as well as the dark current (both Ohmic and not) were completely stable. Increasing the voltage (20-40~mC/cm$^{2}$) the total dark current shows a slight growth, which continued when the voltage was further risen (although in both cases the Ohmic dark current did not increase). Towards the 100~mC/cm$^{2}$ mark (corresponding to full efficiency), a more significant increase of the absorbed current can be observed. Finally, the detector was moved closer to the $^{137}$Cs source and a lower HV was applied to further limit the absorbed current. Nonetheless, the current steadily increased and, this time, the Ohmic dark current increased as well. As a follow-up to these observations, it was decided to investigate whether a significant variation in electrode resistivity was taking place since this could possibly explain the absorbed current variations.

\begin{figure}[h!]
\center \includegraphics[height=0.6\linewidth]{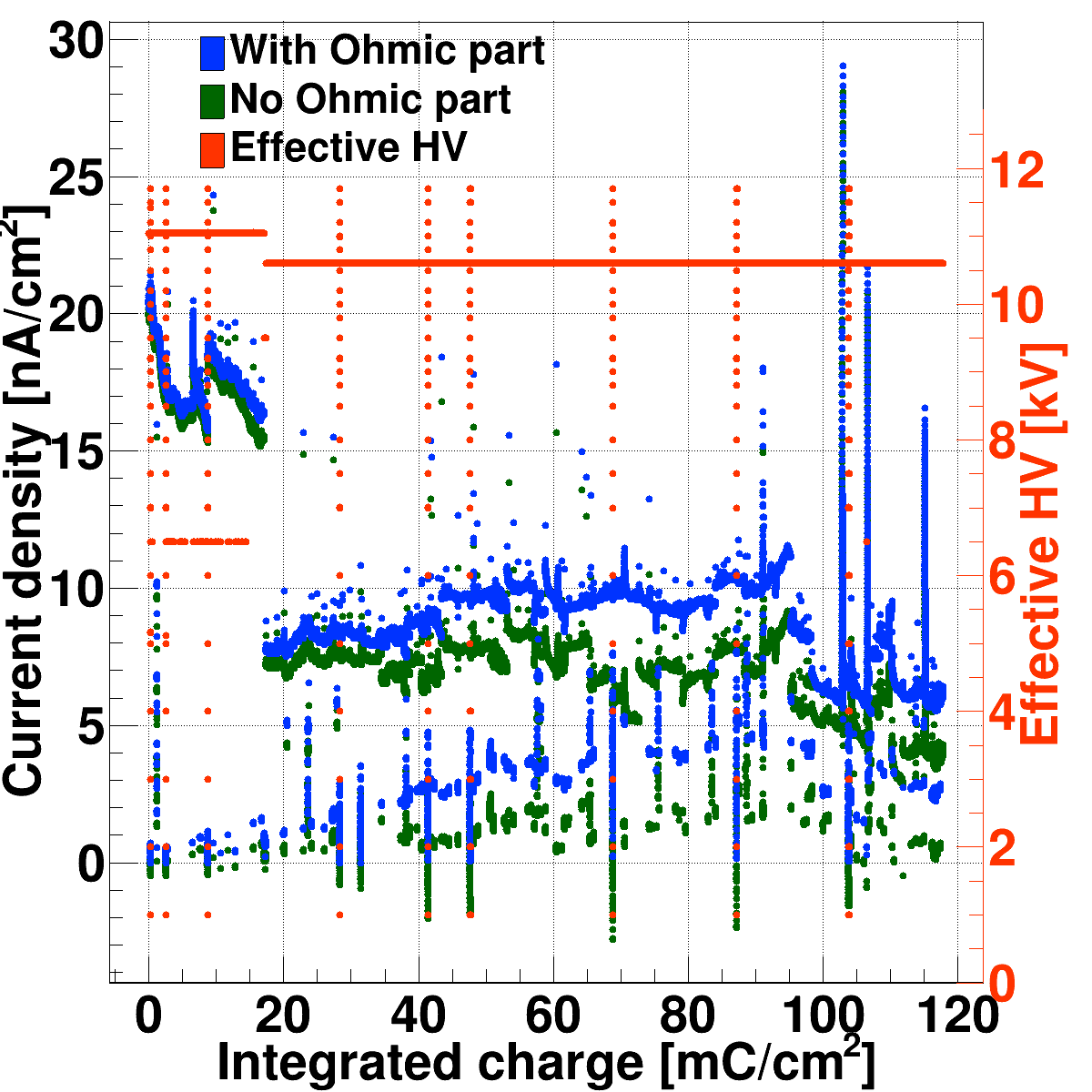}
\caption{Trend of current density and effective high voltage as a function of the integrated charge for the CMS RE11 BOT gap}
\label{fig:cmsAging}
\end{figure}

A somewhat different behavior can be observed by looking at Figure \ref{fig:cmsAging}, which shows the results obtained with the BOT gap of the CMS RE11 detector. 

The increase of separation between the blue and green curves (in the 40-100~mC/cm$^{2}$ range) is a sign of an increase in absorbed Ohmic dark current, which seems to fade after the $\approx$100~mC/cm$^{2}$ mark. The total dark current seems to follow a similar trend. Nonetheless the current for all the three gaps of the CMS RE11 detector shows a more stable behavior over time. 

\begin{figure}[h!]
\center \includegraphics[height=0.6\linewidth]{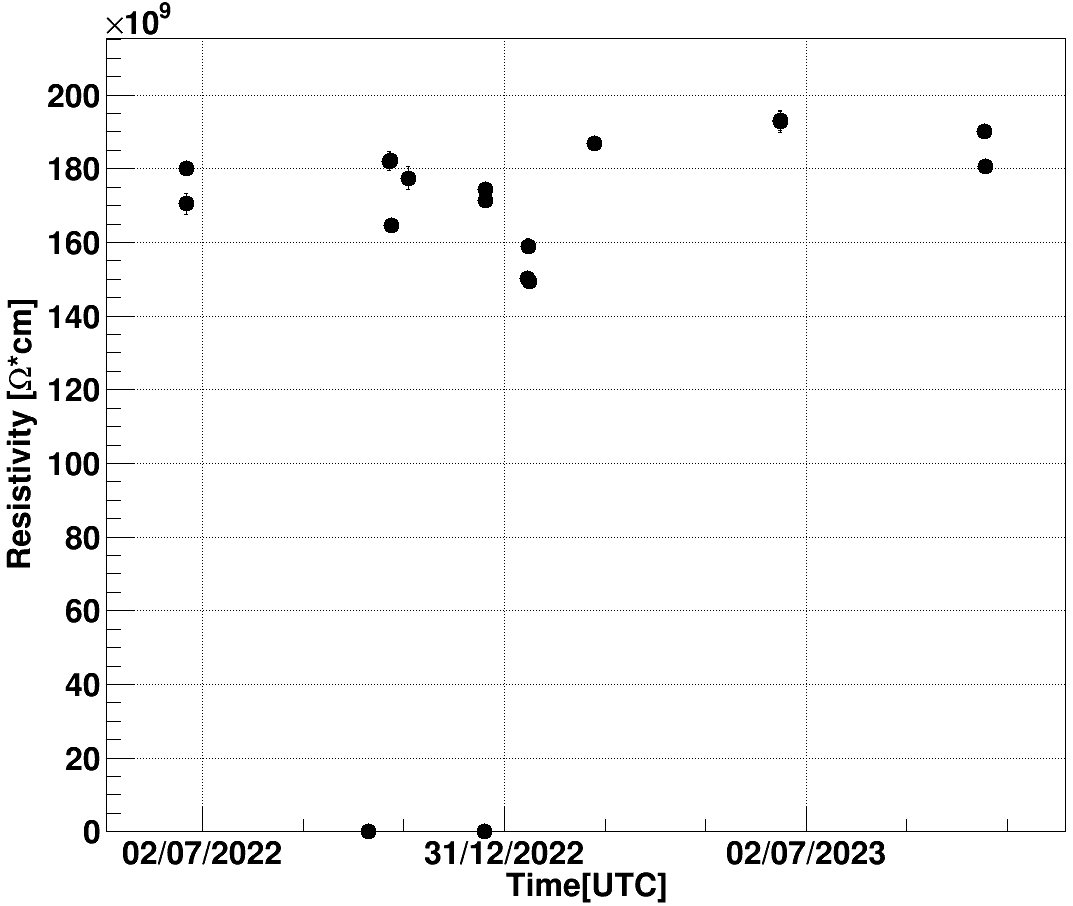}
\caption{Trend of the electrode resistivity (measured with the Ar method) during the aging test time period for the SHiP/LHCb detector (the 0 values represent measurements where the detector was not included)}
\label{fig:res}
\end{figure}

As it was anticipated, a significant change in the resistivity of the electrodes could explain the observed trend. Indeed, using the Ar method (described in \cite{Ar}), one can calculate this quantity and periodic resistivity measurements are being carried out to monitor its evolution. Figure \ref{fig:res} shows the trend of the electrode resistivity over time for the LHCb/SHiP detector. The value seems to be quite stable over time, hence the change in absorbed current cannot be attributed to a simple change in electrode resistivity and a more in-depth investigation is needed to uncover the root cause of this behavior. Similar observations can be made for the BOT gap of the CMS RE11 detector. Indeed, its resistivity shows an increasing trend over time and this cannot explain the observed behavior of the absorbed current.

\begin{figure}
\begin{subfigure}[h]{0.49\linewidth}
\includegraphics[width=\linewidth]{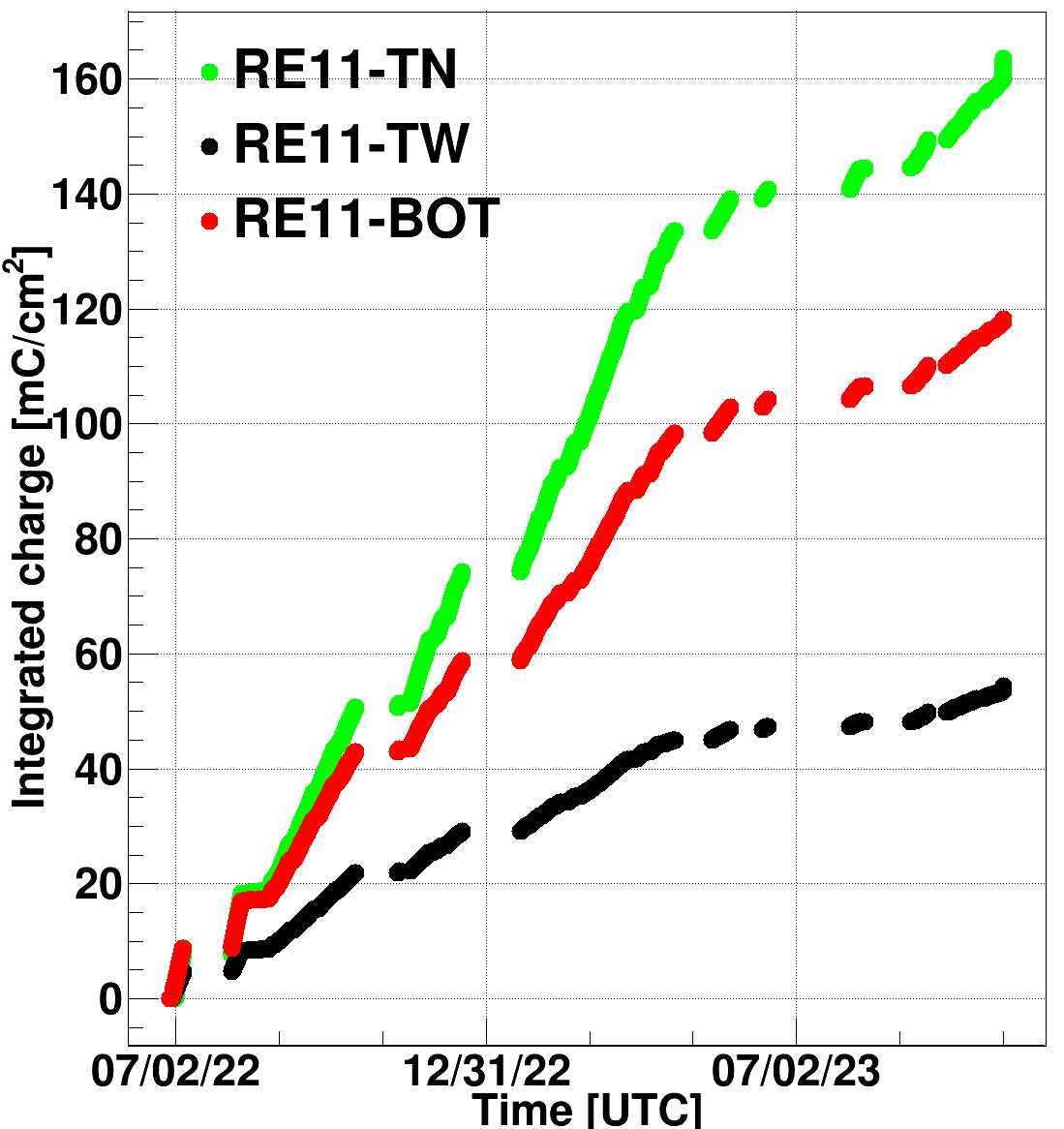}
\caption{Image A}
\label{fig:agingTot1}
\end{subfigure}
\hfill
\begin{subfigure}[h]{0.49\linewidth}
\includegraphics[width=\linewidth]{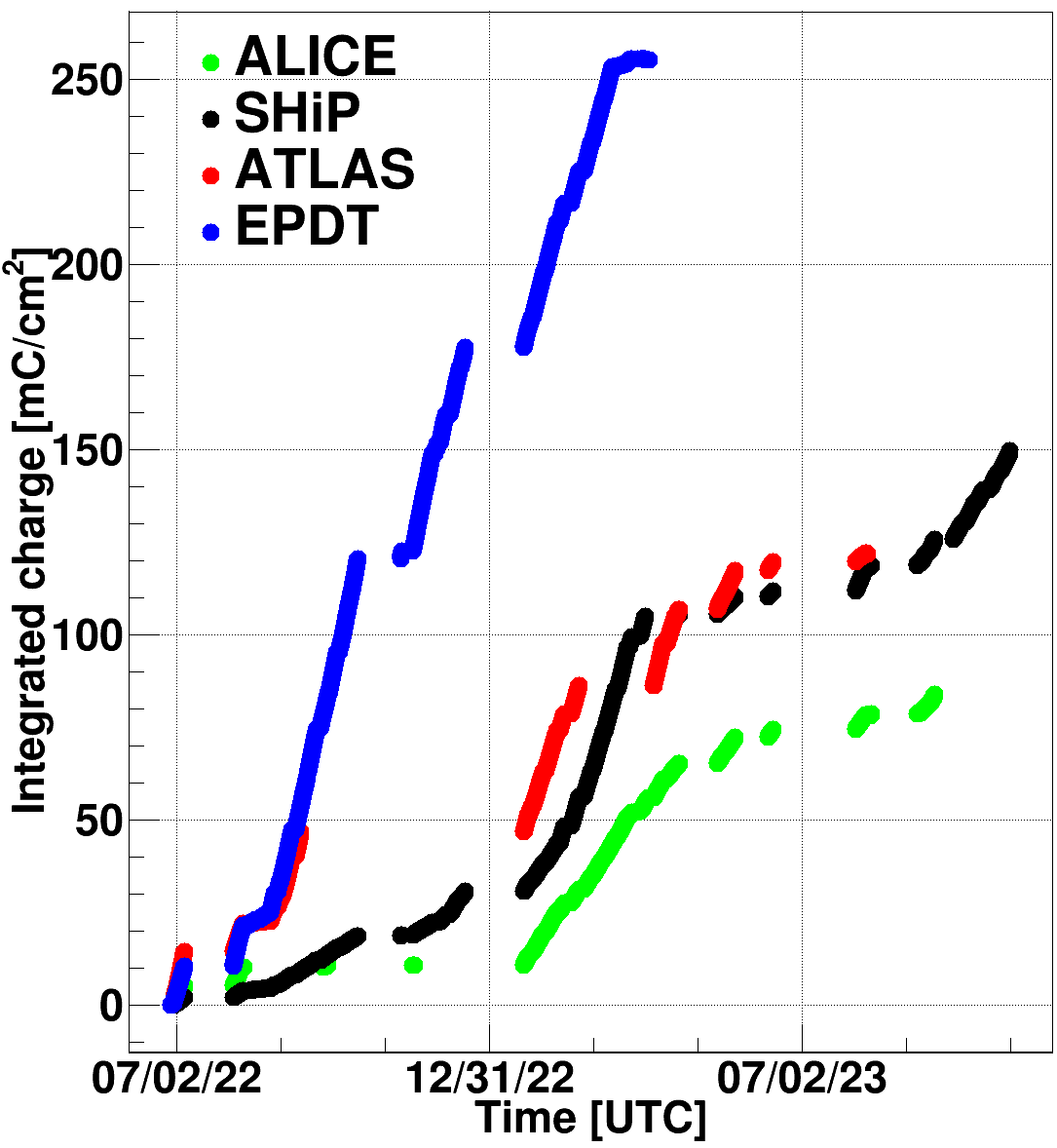}
\caption{Image B}
\label{fig:agingTot2}
\end{subfigure}%
\caption{This is a figure with two subfigures}
\end{figure}

Figures \ref{fig:agingTot1} and \ref{fig:agingTot2} show the trend of the integrated charge over time for all the detectors of the collaboration. The latter figure refers to the three gaps of the CMS RE11 detector while the former to all the other RPCs. On average, a total of $\approx$100~mC/cm$^{2}$ has been accumulated by all the RPCs (the slight difference among them can be attributed to the different efficiency at the irradiation voltage as well as different distances from the irradiation source).

\begin{figure}[h!]
\center \includegraphics[width=\linewidth]{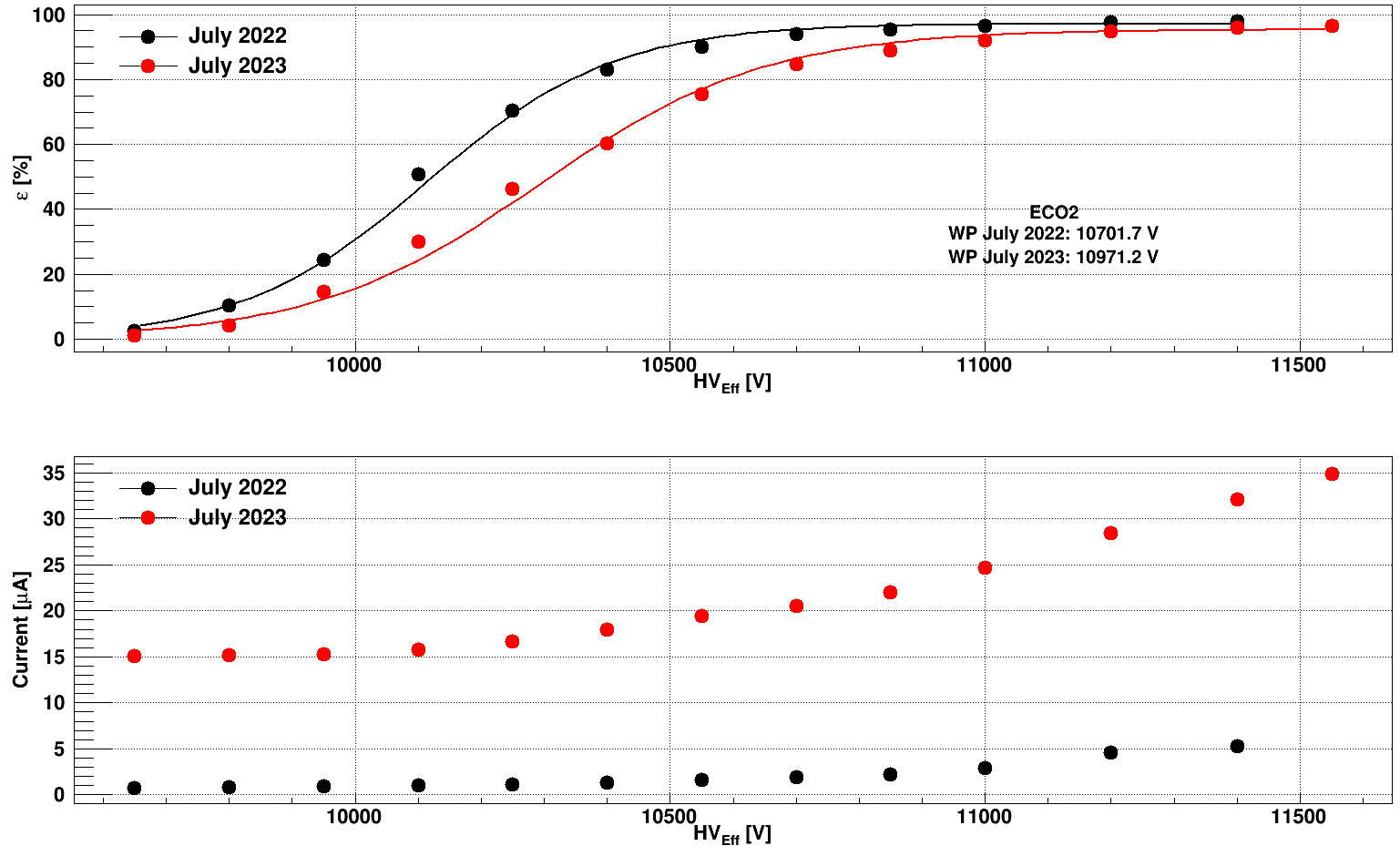}
\caption{Top panel: comparison of source-off efficiency curves between July 2022 and 2023, obtained with the ECO2 gas mixture. Bottom panel: comparison of dark current curves between July 2022 and 2023, obtained with the ECO2 gas mixture}
\label{fig:comparison}
\end{figure}

In July 2023 a second beam test campaign was carried out, in order to compare the RPC performance to the baseline values obtained in July 2022. This comparison has, so far, only been carried out for the ALICE RPC and the preliminary results are shown in Figure \ref{fig:comparison}. The top panel of said figure shows the comparison between the source-off efficiency in July 2022 (black curve) and July 2023 (red curve) while the bottom panel shows (with the same color convention of the top panel) the dark current measured in the two beam tests.

The efficiency curves have been interpolated using a logistic function (as explained in \cite{focusPoint}) to extract the working point, defined as the high voltage knee (HV$_{eff}$ where the efficiency is 95\% of its plateau value) + 150~V. The comparison of the two curves shows no appreciable decrease of the maximum efficiency, although the working point is shifted to higher voltages by $\approx$250~V while the slope of the 2023 curve seems to be decreasing, with respect to the 2022 data. The comparison of the dark current curves shows an increase between the two beam tests and further studies are ongoing to understand if the working point shift could be justified by the observed current increase. Moreover, the analyses of the data from other collaboration members will show if similar effects are observed in other detectors as well.

\vspace{-25pt}

\section{Conclusions and outlook}
\label{sec:conclusion}

\vspace{-8pt}

The search for an eco-friendly alternative gas mixture for RPC detectors is a hot-topic in the gaseous detector community, especially due to the new EU regulations which foresee a progressive phase down in the production and usage of F-gases (which represent more than 90\% of the currently employed gas mixture).

Several eco-friendly alternatives, where C$_{2}$H$_{2}$F$_{4}$ is replaced by different concentrations of HFO/CO$_{2}$ have been identified using cosmic muons and, currently, a long-term aging campaing is ongoing. Two different eco-friendly alternatives (ECO1 and ECO2) have been tested. For both gas mixtures the detector working point is $\approx$2 and 1~kV higher than the one of the currently employed gas mixture respectively.

ECO1 was discarded due to current instabilities in all detectors (observed after the integration of only $\approx$10~mC/cm$^{2}$) as well as due to its higher working point, possibly leading to a greater production of impurities. ECO2 has shown a less unstable behavior, for some detectors more than for others and more in depth analysis (and hardware upgrades) are ongoing to clarify this observation. In general, a total of $\approx$100~mC/cm$^{2}$ has been integrated up to now by all the detectors of the ECOgas collaboration.

A preliminary comparison of the RPC performance before and after the irradiation campaign (using the muon beam) has been carried out for the ALICE detector and it has shown that the maximum efficiency reached has not appreciably decreased, although a shift of the WP of $\approx$150-250~V (depending on the mixture) has been observed and further analysis is ongoing to possibly correlate it with the observed increase in absorbed current.

\end{document}